

\input harvmac

\input amssym.def
\input amssym
\baselineskip 14pt
\magnification\magstep1
\parskip 6pt

\font \bigbf=cmbx10 scaled \magstep1

\newdimen\itemindent \itemindent=32pt
\def\textindent#1{\parindent=\itemindent\let\par=\resetpar%
\indent\llap{#1\enspace}\ignorespaces}

\let\oldpar=\par
\def\resetpar{\oldpar\parindent=20pt\let\par=\oldpar}

\font\ninerm=cmr9 \font\ninesy=cmsy9
\font\eightrm=cmr8 \font\sixrm=cmr6
\font\eighti=cmmi8 \font\sixi=cmmi6
\font\eightsy=cmsy8 \font\sixsy=cmsy6
\font\eightbf=cmbx8 \font\sixbf=cmbx6
\font\eightit=cmti8
\def\eightpoint{\def\rm{\fam0\eightrm}
  \textfont0=\eightrm \scriptfont0=\sixrm \scriptscriptfont0=\fiverm
  \textfont1=\eighti  \scriptfont1=\sixi  \scriptscriptfont1=\fivei
  \textfont2=\eightsy \scriptfont2=\sixsy \scriptscriptfont2=\fivesy
  \textfont3=\tenex   \scriptfont3=\tenex \scriptscriptfont3=\tenex
  \textfont\itfam=\eightit  \def\it{\fam\itfam\eightit}%
  \textfont\bffam=\eightbf  \scriptfont\bffam=\sixbf
  \scriptscriptfont\bffam=\fivebf  \def\bf{\fam\bffam\eightbf}%
  \normalbaselineskip=9pt
  \setbox\strutbox=\hbox{\vrule height7pt depth2pt width0pt}%
  \let\big=\eightbig  \normalbaselines\rm}
\catcode`@=11 %
\def\eightbig#1{{\hbox{$\textfont0=\ninerm\textfont2=\ninesy
  \left#1\vbox to6.5pt{}\right.\n@@space$}}}
\def\vfootnote#1{\insert\footins\bgroup\eightpoint
  \interlinepenalty=\interfootnotelinepenalty
  \splittopskip=\ht\strutbox %
  \splitmaxdepth=\dp\strutbox %
  \leftskip=0pt \rightskip=0pt \spaceskip=0pt \xspaceskip=0pt
  \textindent{#1}\footstrut\futurelet\next\fo@t}
\catcode`@=12 %

\def\a{\alpha}
\def\b{\beta}
\def\c{\gamma}
\def\d{\delta}

\def\l{\lambda}
\def\m{\mu}
\def\n{\nu}

\def\r{\rho}

\def\C{\Gamma}

\def\L{\Lambda}

\def\pl{\partial}

\lref\DH{I.T. Drummond and S. Hathrell, Phys. Rev. D22 (1980) 343 }
\lref\Sone{R.D. Daniels and G.M. Shore, Nucl. Phys. B425 (1994) 634 }
\lref\Stwo{R.D. Daniels and G.M. Shore, Phys. Lett. B367 (1996) 75 }
\lref\Sthree{G.M. Shore, Nucl. Phys. B460 (1996) 379 }
\lref\Bondi{H. Bondi, M.G.J. van der Burg and A.W.K. Metzner, Proc. Roy. Soc.
A269 (1962) 21}
\lref\Sachs{R.K. Sachs, Proc. Roy. Soc. A270 (1962) 103 }
\lref\Done{I.T. Drummond, private communication}
\lref\Dtwo{I.T. Drummond, gr-qc/9908058} 
\lref\ABJone{J.D. Barrow and J. Magueijo, Class. Quant. Grav. 16 (1999) 1435} 
\lref\ABJtwo{A. Albrecht and J. Magueijo, Phys. Rev. D59 (1999) 043516 }
\lref\ABJthree{J. Magueijo, Phys. Rev. D62 (2000) 103521 } 
\lref\DN{A.D. Dolgov and I.D. Novikov, Phys. Lett. B442 (1998) 82 }
\lref\Ch{S. Chandresekhar, {\it The Mathematical Theory of Black Holes},
Clarendon, Oxford (1985) }
\lref\Inverno{R.A. d'Inverno, {\it Introducing Einstein's Relativity},
Clarendon, Oxford (1992) }


{\nopagenumbers
\rightline{CERN-TH/2000-285}
\rightline{SWAT/266}
\rightline{UGVA-DPT-00-9-1088 }
\rightline{gr-qc/0012XXX}
\vskip1cm
\centerline{\bigbf Accelerating Photons with Gravitational Radiation}

\vskip1cm

\centerline {\bf G.M. Shore\foot{
Permanent address:~~Department of Physics, University of Wales Swansea, 
Singleton Park, Swansea, SA2 8PP, U.K.}
}
\vskip0.5cm
\centerline{\it D\'epartement de Physique Th\'eorique,}
\centerline{\it Universit\'e de Gen\`eve}
\centerline{\it 24, quai E.~Ansermet,}
\centerline{\it CH 1211 Geneva 4, Switzerland}
\vskip0.2cm
\centerline{and}
\vskip0.2cm
\centerline{\it TH Division, CERN,}
\centerline{\it CH 1211 Geneva 23, Switzerland}

\vskip1cm

{
\parindent 1.5cm{

{\narrower\smallskip\parindent 0pt
The nature of superluminal photon propagation in the gravitational field
describing radiation from a time-dependent, isolated source (the Bondi-Sachs 
metric) is considered in an effective theory which includes interactions which
violate the strong equivalence principle. Such interactions are, for example, 
generated by vacuum polarisation in conventional QED in curved spacetime.
The relation of the resulting light-cone modifications to the Peeling Theorem 
for the Bondi-Sachs spacetime is explained. 
\narrower}}}

\vskip2cm

\leftline{CERN-TH/2000-285}
\leftline{SWAT/266} 
\leftline{UGVA-DPT-00-9-1088 }
\leftline{December 2000}

\vfill\eject}

\pageno=1

\newsec{Introduction}

The possibility of superluminal photon propagation in gravitational fields is 
an intriguing prediction of quantum field theory in curved spacetime.
It raises fundamental questions about the realisation of causality
and is potentially of great importance for early-universe cosmology. 

In the context considered here, superluminal propagation was originally
discussed in a pioneering paper by Drummond and Hathrell \refs{\DH}, who 
investigated
the effects of vacuum polarisation on photon propagation in a variety of
classical curved spacetimes, including Schwarzschild, Robertson-Walker and
weak-field gravitational waves. They showed that in general it is possible
to find directions and polarisations for which the photon velocity exceeds
the fundamental constant $c$. This work has subsequently been generalised to 
other examples of background spacetimes, including Reissner-Nordstrom and 
Kerr black holes \refs{\Sone,\Stwo}. 
In ref.\refs{\Sthree}, we presented some further theoretical
analysis, including the formulation of a polarisation sum rule and a
horizon theorem, based on the original work of \refs{\DH}. In this paper, we extend
this development and present results for a phenomenologically important
spacetime, the Bondi-Sachs metric describing gravitational radiation from an
isolated source \refs{\Bondi,\Sachs}.

The essential physics underlying the Drummond-Hathrell mechanism for
superluminal propagation is a violation of the strong equivalence principle.
Here, we understand the weak equivalence principle (WEP) to be
the requirement that spacetime is Riemannian and thus has at each point
a local inertial frame (LIF). By the strong equivalence principle
(SEP), we mean the further assumption that the laws of physics are the same 
in the LIFs at each point of spacetime, and take their special relativistic 
form at the origin of each LIF. In particular, this condition states that
matter couples to the gravitational field only via the connection, 
with no direct curvature coupling. It is clear that the two forms
of the equivalence principle have a quite different status. While the
WEP is fundamental to the structure of general relativity, the SEP appears to be 
merely an extra dynamical assumption (minimal coupling) which may not be 
essential for the self-consistency of the theory. Of course, this is precisely 
what we are testing by studying superluminal propagation and causality in a
situation where the SEP is relaxed. If the theory is indeed still 
self-consistent, it then becomes an experimental question whether or not 
SEP-violating interactions exist and with what strength. This paper is concerned 
with the characteristics of photon propagation in
a particularly relevant gravitational field, that created by a time-dependent,
isolated, radiating source. 
 
The particular SEP-violating interactions we consider here are given by
the effective action
\eqn\action{
\C = \int dx \sqrt{-g}\biggl[
-{1\over4}F_{\m\n}F^{\m\n} + {1\over m^2}\biggl(
a R F_{\m\n}F^{\m\n} + b R_{\m\n} F^{\m\l} F^\n{}_\l
+ c R_{\m\l\n\r} F^{\m\l} F^{\n\r} \biggr)\biggr]
}
As shown by Drummond and Hathrell, this action arises through
vacuum polarisation effects at one-loop level in QED in curved spacetime.
In this case, $a = -{1\over144}{\a\over\pi}$, $b = {13\over360}{\a\over\pi}$ 
and $c = -{1\over360}{\a\over\pi}$ and the scale $m$ is the electron mass.
Further corrections to the effective action involving higher derivatives
of the field strengths and curvatures also arise in QED and are relevant to 
the question of dispersion and high-frequency propagation, but these will
not be considered here. Similar effective actions may also be expected to
arise generically as low-energy approximations to more speculative
fundamental theories of quantum gravity. However, at this point we do not need
to restrict ourselves to any particular mechanism, but can instead consider
the effective action \action ~as a phenomenological model involving a new 
fundamental scale $m$ to be determined by experiment. This is the approach we
will adopt in this paper.

As reviewed in the next section, this action induces curvature-dependent
modifications to the effective light cone governing photon propagation
and leads to the possibility of superluminal velocities.
Two important generic features were formalised in ref.\refs{\Sthree}.
First, for Ricci flat spacetimes there exists a polarisation sum rule
whereby if one polarisation state propagates with velocity less than $c$,
the other must necessarily have a velocity greater than $c$.
Second, for black hole spacetimes, even if the photon velocity along some 
trajectory is different from $c$, it reverts to the standard light cone
velocity precisely on the event horizon, ensuring that the physical and
geometrical horizons coincide.
These features are most apparent in the following formula for the effective 
light cone, following from eq.\action:
\eqn\lightcone{
k^2 = -{16\pi\over m^2}(b+2c) T_{\m\n}k^\m k^\n 
+ {8c\over m^2} C_{\m\l\n\r}k^\m k^\n a^\l a^\r
}
where $k^\m$ is the wave vector (photon momentum) and $a^\l$ is the
polarisation. In this form, we have used the Einstein equations (with $G=1$) 
to substitute the energy-momentum tensor $T_{\m\n}$ for the Ricci tensor.
The first term is polarisation-independent and is proportional to the
projection of the energy-momentum tensor appearing in the weak energy
condition, viz. $T_{\m\n}k^\m k^\n \ge 0$ for any null vector $k^\m$.
A similar modification to $k^2$ also arises for photon propagation
in other modified environments such as background electromagnetic fields
or finite temperature. The second contribution, 
which is special to gravitational backgrounds, produces a polarisation
dependent shift in the effective light cone proportional to a particular 
projection of the Weyl tensor (i.e.~a particular Newman-Penrose scalar, as 
explained in section 3). The shift is equal and opposite for the two 
polarisations, one of which is therefore necessarily superluminal.

A polarisation-dependent shift in the light cone (gravitational birefringence)
allowing superluminal propagation is therefore the essential signature
of SEP-violating interactions in curved spacetime electrodynamics.

If this effect is indeed observable, there are a number of potentially
important cases which should be considered. The first, already discussed
in ref.\refs{\DH}, is the polarisation dependence of the bending of light by 
a star or other matter distribution whose exterior field is described by the 
Schwarzschild metric. In this case, the angle of deflection becomes dependent
on the photon polarisation.  Superluminal propagation would be especially
important in the evolution of the early universe, where it could be relevant
for the horizon problem \refs{\Done, \ABJone, \ABJtwo, \ABJthree}. 
This probably requires changes in $c$ by many 
orders of magnitude, rather than the perturbative corrections considered 
here (we are implicitly considering \action ~to be the leading terms in 
an expansion in $O(R/m^2)$). Nevertheless, \lightcone ~does predict 
superluminal propagation for FRW spacetimes \refs{\DH}. Here, the spatial 
isotropy ensures that the second term vanishes ($C_{\m\l\n\r} = 0$
for FRW spacetime), but the first term produces a polarisation-independent
shift in the light cone which implies superluminal velocities provided
the weak energy condition $\r + p \ge 0$ is satisfied.
On the other hand, for de Sitter spacetime, $k^2$ remains zero because 
the spacetime is isotropic.

Here, we consider a third important case: the propagation of light in the 
background spacetime corresponding to a time-dependent, isolated source
emitting gravitational radiation. We show that there is a 
polarisation-dependent superluminal effect and express this in terms of the
appropriate Newman-Penrose scalars characterising the Weyl tensor.
The asymptotic behaviour depends on the direction of the photons. For
ingoing photons, the light cone shift is of $O(1/r)$ and is given by the
amplitude of the gravitational waves far from the source; for the more 
important case of outgoing photons, the shift is $O(1/r^5)$ and depends
on the `quadrupole aspect' of the Bondi-Sachs metric. These results are
related to the Peeling Theorem \refs{\Sachs}, 
which describes the asymptotic dependence
of the Newman-Penrose scalars in the Bondi-Sachs metric. 

Before deriving these results, we mention briefly two interesting recent
developments in the theory of superluminal propagation. First, Albrecht, 
Barrow and Magueijo \refs{\ABJone, \ABJtwo, \ABJthree} 
have studied the implications for the early universe
of allowing the fundamental constant $c$ to be time-dependent. Although this
formalism is very different from that considered here, these papers illustrate
well the potential importance of superluminal effects in cosmology.
Second, Drummond \refs{\Dtwo} has recently proposed a manifestly covariant 
bi-metric 
theory of gravity in which matter and gravity couple to different vierbeins
whose relative orientation is determined dynamically through a sigma model 
action. He then applies this theory to the specific problem of `dark matter',
arguing that the phenomena usually attributed to the existence of dark matter
can alternatively be described by a modification of the fundamental theory
of gravity. This is of course a much more radical generalisation of
conventional gravity than considered in this paper, but this new theory would
certainly incorporate the type of effects discussed here.

In the rest of the paper, we review briefly the derivation of the
modified light cone condition \lightcone ~in section 2. Then, in section 3, 
we describe the Bondi-Sachs metric and the Peeling Theorem. The results 
on photon propagation and our general conclusions are given in 
sections 4 and 5.

\newsec{Superluminal Propagation in Gravitational Fields}

The characteristics of photon propagation following from the effective
action \action ~are most easily described using geometric optics \refs{\DH}.
In this picture, the electromagnetic field is written as the product
of a slowly-varying amplitude and a rapidly-varying phase, i.e.
\eqn\sectba{
F_{\m\n} = f_{\m\n}e^{i\vartheta}
}
The wave vector (photon momentum) is defined as $k_\m = \pl_\m \vartheta$,
while the Bianchi identity constrains $f_{\m\n}$ to have the form
\eqn\sectbb{
f_{\m\n} = k_\m a_\n - k_\n a_\m
}
where the direction of $a^\m$ specifies the polarisation. For physical
polarisations, $k_\m a^\m = 0$.

For conventional curved spacetime QED based on the usual Maxwell action,
the equation of motion is simply 
\eqn\sectbc{
D_\m F^{\m\n} = 0~~~~~~~~~~~
\Rightarrow  ~~~~~~k_\m f^{\m\n} = 0
}
Since this implies
\eqn\sectbcc{
k^2 a^\n = 0
}
we immediately deduce that $k^2=0$, i.e.~$k^\m$ is a null vector.
Then, from its definition as a gradient, we see 
\eqn\sectbd{
k^\m D_\m k^\n ~=~ k^\m D^\n k_\m ~=~ {1\over2} D^\n k^2 ~=~ 0
}
Light rays (photon trajectories) are defined as the integral curves of the
wave vector, i.e.~the curves $x^\m(s)$ where ${dx^\m \over ds} = k^\m$.
These curves therefore satisfy
\eqn\sectbe{
0 ~=~ k^\m D_\m k^\n 
~=~ {d^2 x^\n \over ds^2} + \C^\n_{\m\l} {dx^\m \over ds}{dx^\l \over ds}
}
which is the geodesic equation. So in the conventional theory, light rays
are null geodesics.

The effective action \action ~gives rise to a modified equation of motion 
which, under the approximations listed below, implies
\eqn\kf{
k_\m f^{\m\n} ~+~ {1\over m^2} \biggl[2b R^\m{}_\l k_\m f^{\l\n}
+ 4c R^{\m\n}{}_{\l\r} k_\m f^{\l\r} \biggr] ~~=~~0
}
Here, we have made the standard geometric optics approximation of
neglecting derivatives of $f_{\m\n}$ relative to the derivatives of the 
phase factor, which produce powers of the momentum $k_\m$; we have neglected
derivatives of the curvature tensor, which would  produce corrections of
$O(\l/L)$, where $\l$ is the photon wavelength and $L$ is a typical curvature 
scale; and we have omitted the corrections to the term $D_\m F^{\m\n}$ 
of $O(aR/m^2)$ coming from the Ricci scalar in the effective action -- these
are of $O(\l_c^2/L^2)$, where $\l_c$ is the Compton wavelength corresponding
to a particle of mass $m$ (the electron in the conventional QED derivation
of \action ~) which should be neglected as higher order if we view \action ~as
the leading term in an expansion in powers of curvature.

Eq.\kf ~can now be rewritten as an equation for the polarisation vector
$a^\m$, and re-expressing in terms of the Weyl tensor we find
\eqn\ka{
k^2 a^\n ~+~ {(2b+4c)\over m^2}~ R_{\m\l} \bigl(k^\m k^\l a^\n
- k^\m k^\n a^\l \bigr) ~+~ {8c\over m^2} ~ 
C_\m{}^\n{}_{\l\r} k^\m k^\l a^\r ~~=~~0
}
The solutions of this equation describe the characteristics of propagation
for a photon of momentum $k^\m$ and polarisation $a^\m$. Contracting
with $a^\m$ (and assuming spacelike normalisation $a^\m a_\m = -1$),
we find the effective light cone
\eqn\ksquare{
k^2 ~+~{(2b+4c)\over m^2} R_{\m\l} k^\m k^\l ~-~{8c\over m^2} 
C_{\m\n\l\r} k^\m k^\l a^\n a^\r ~~=~~0
}
from which \lightcone ~follows immediately.

It should be noted that all these equations are manifestly local
Lorentz invariant. On the other hand, the presence of the explicit
curvature coupling in the effective action means that the equations 
of motion do not reduce to their special relativistic form at the origin 
of each LIF, and thus that the dynamics is different in the LIFs at
different points in spacetime. In this sense, these equations 
violate the strong principle of equivalence. Some implications of this
for causality have been discussed in refs.\refs{\DH,\Sthree,\DN}.

At this point, it is illuminating to re-write the effective light cone
condition using the Newman-Penrose tetrad formalism, and in particular to
show the dependence on the NP scalars characterising the Weyl tensor.
The first step is to choose a null tetrad as follows. Let $\ell^\m$
be a null vector. 
Let $a^\m$ and $b^\m$ be spacelike, transverse 
vectors and define the complex null vectors $m^\m$ and $\bar m^\m$
by $m^\m = {1\over\sqrt2}(a^\m + i b^\m)$ and $\bar m^\m = {1\over\sqrt2}
(a^\m - i b^\m)$. Finally, choose a further null vector $n^\m$ orthogonal
to $m^\m$ and $\bar m^\m$. These vectors satisfy the orthogonality
conditions:
\eqn\sectbf{
\ell.m ~=~ \ell.\bar m ~=~ n.m ~=~ n.\bar m  ~=~ 0
}
they are null vectors:
\eqn\sectbg{
\ell.\ell ~=~ n.n ~=~ m.m ~=~ \bar m . \bar m ~=~ 0
}
and we impose the normalisation conditions:
\eqn\sectbh{
\ell.n ~=~1  ~~~~~~~~~~~~~~~~~~~~~~~ m.\bar m ~=~ -1
}
The null tetrad is defined by the vierbeins $e_a^\m$, where we
define $e_1^\m = \ell^\m$, $e_2^\m = n^\m$, $e_3^\m = m^\m$
and $e_4^\m = \bar m^\m$. The corresponding metric is
\eqn\sectbi{
\eta_{ab} ~=~ \left(\matrix{0&1&0&0\cr 1&0&0&0\cr
0&0&0&-1\cr 0&0&-1&0\cr}\right)
}

The five complex NP scalars characterising the Weyl tensor are
(following closely the notation of ref.\refs{\Ch}) 
\eqnn\sectbone
$$\eqalignno{
\Psi_0 ~&=~ - C_{abcd}\ell^a m^b \ell^c m^d ~~~&=~~
-C_{1313} ~~~~~~~~~~~~~~~~~~~~~ \cr
\Psi_1 ~&=~ - C_{abcd}\ell^a n^b \ell^c m^d ~~~&=~~
-C_{1213} ~~~~~~~~~~~~~~~~~~~~~ \cr
\Psi_2 ~&=~ - C_{abcd}\ell^a m^b \bar m^c n^d ~~~&=~~
-C_{1342} ~~~~~~~~~~~~~~~~~~~~~ \cr
\Psi_3 ~&=~ - C_{abcd}\ell^a n^b \bar m^c n^d ~~~&=~~
-C_{1242} ~~~~~~~~~~~~~~~~~~~~~ \cr
\Psi_4 ~&=~ - C_{abcd}n^a \bar m^b n^c \bar m^d ~~~&=~~
-C_{2424} ~~~~~~~~~~~~~~~~~~~~~ \cr
{}&{}& \sectbone \cr }
$$
The symmetries of the Weyl tensor imply several interesting relations 
amongst its components. Most important for our discussion are the 
trace-free conditions
\eqn\sectbj{
\eta^{ad} C_{abcd} = 0
}
and cyclicity, e.g.
\eqn\sectbk{
C_{1234} + C_{1342} + C_{1423} = 0
}
Together, these imply the important identity
\eqn\weylid{
C_{\m\n\l\r} \ell^\m m^\n \ell^\l \bar m^\r ~~=~~C_{1314} ~~=~~0
}

Components of the Ricci tensor have a similar classification.
We define
\eqnn\sectbtwo
$$\eqalignno{
\Phi_{00} ~&=~ -{1\over2} R_{\m\n}\ell^\m \ell^\n ~=~ 
- {1\over2} R_{11} ~~~~~~~~~~~~~~~~~~
&\Phi_{22} ~=~ -{1\over2} R_{\m\n}n^\m n^\n ~=~ - {1\over2} R_{22} \cr
\Phi_{02} ~&=~ -{1\over2} R_{\m\n}m^\m m^\n ~=~ - {1\over2} R_{33} 
&\Phi_{20} ~=~ -{1\over2} R_{\m\n}\bar m^\m \bar m^\n ~=~ 
- {1\over2} R_{44} \cr
\Phi_{01} ~&=~ -{1\over2} R_{\m\n}\ell^\m m^\n ~=~ 
- {1\over2} R_{13} ~~~~~~~~~~~~~~~~~~
&\Phi_{10} ~=~ -{1\over2} R_{\m\n}\ell^\m \bar m^\n ~=~ 
- {1\over2} R_{14} \cr
\Phi_{12} ~&=~ -{1\over2} R_{\m\n}n^\m m^\n ~=~ - {1\over2} R_{23} 
&\Phi_{21} ~=~ -{1\over2} R_{\m\n}n^\m \bar m^\n ~=~ 
- {1\over2} R_{24} \cr
&\Phi_{11} ~=~ -{1\over4} (R_{\m\n}\ell^\m n^\n  + R_{\m\n}m^\m \bar m^\n)~=~ 
- {1\over4} (R_{12} + R_{34}) &{} \cr
&\L ~=~ {1\over24} R ~=~
{1\over12} (R_{\m\n}\ell^\m n^\n  - R_{\m\n}m^\m \bar m^\n)~=~
{1\over12} (R_{12} - R_{34}) &{} \cr
{}&{}&\sectbtwo\cr}
$$

We can now re-express the light cone condition in NP form, using the identity
\weylid . 
For example, if we choose the (unperturbed) photon momentum in the direction 
of the null vector $\ell^\m$, i.e. $k^\m = \omega\ell^\m$, and the transverse 
polarisation vectors as $a^\m$, $b^\m$, then the term in
\ksquare ~proportional to the Weyl tensor becomes, for the two 
polarisations respectively,
\eqnn\sectbthree
$$\eqalignno{
\pm {1\over2} C_{\m\n\l\r} \ell^\m \ell^\l 
\bigl(m^\n \pm \bar m^\n\bigr) \bigl(m^\r \pm \bar m^\r \bigr)~~&=~~
\pm {1\over2} \bigl( C_{\m\n\l\r} \ell^\m m^\n \ell^\l m^\r 
+ C_{\m\n\l\r} \ell^\m \bar m^\n \ell^\l \bar m^\r \bigr) \cr 
{}&=~~ \pm {1\over2} (\Psi_0 + \Psi_0^*) &\sectbthree\cr}
$$
The lightcone condition is therefore simply
\eqn\lightconenew{
k^2 ~~=~~ {(4b+8c)\omega^2\over m^2} \Phi_{00} ~~\pm~~
{4c\omega^2\over m^2} (\Psi_0 + \Psi_0^*)
}
depending on the polarisation. It is interesting that this depends 
on only a single NP scalar for each of the Ricci and Weyl tensor 
contributions. The polarisation sum rule and horizon theorem described 
in ref.\refs{\Sthree} are immediate consequences of this form of the 
lightcone condition. Other choices of the photon momentum and
polarisation give analogous expressions for the modified light cone.

In the rest of this paper, we apply this result to the special example 
of the Bondi-Sachs metric and show the precise relation to the 
Peeling Theorem.

\vfill\eject

\newsec{Bondi-Sachs Metric and the Peeling Theorem}

The spacetime describing an isolated, radiating source is given by the
Bondi-Sachs metric:
\eqn\sectca{
ds^2 = -W du^2 - 2 e^{2\b} du dr + r^2 h_{ij}(dx^i - U^i du)
(dx^j - U^j du)
}
where
\eqn\sectcb{
h_{ij}dx^i dx^j = {1\over2}(e^{2\c} + e^{2\d}) d\theta^2
+ 2 \sinh(\c - \d) \sin\theta d\theta d\phi
+ {1\over2}(e^{-2\c} + e^{-2\d}) \sin^2\theta d\phi^2
}
The metric is valid in the vicinity of future null infinity ${\cal I}^+$.
The family of hypersurfaces $u = const$ are null, i.e. $g^{\m\n}
\pl_\m u \pl_\n u = 0$. Their normal vector $\ell_\m$ satisfies
\eqn\sectcc{
\ell_\m = \pl_\m u ~~~~~~~~~~~~~\Rightarrow~~~~~
\ell^2 = 0, ~~~~~~~~\ell^\m D_\mu \ell^\n = 0
}
The curves with tangent vector $\ell^\m$ are therefore
null geodesics; the coordinate $r$ is a radial parameter along these rays  
and is identified as the luminosity distance.

The six independent functions characterising the metric have the following 
asymptotic expansions near ${\cal I}^+$ for large $r$:
\eqnn\sectcone
$$\eqalignno{
W &= 1 - {2{\cal M}\over r} + O({1\over r^2}) \cr
\beta &= -{1\over4} (c_+^2 + c_\times^2) {1\over r^2} + O({1\over r^3}) \cr
{1\over2} (\c + \d) &= {c_+ \over r}
+ {q_+ \over r^3}+ O({1\over r^4})\cr
{1\over2} (\c - \d) &= {c_\times \over r}
+ {q_\times \over r^3}+ O({1\over r^4})\cr
U^\theta + i\sin\theta U^\phi &= 
- {1\over \sin^2\theta} \Bigl(\pl_\theta -
{i\over\sin\theta}\pl_\phi\Bigr) \Bigl(\sin^2\theta(
c_+ + i c_\times)\Bigr){1\over r^2} \cr
{}&~~~~~~~~~~~~~~~~~~~~~~~~~~+ 2\Bigl(d^\theta + i\sin\theta d^\phi + 
\ldots \Bigr){1\over r^3}
+ O({1\over r^4})      &\sectcone\cr
}
$$
where ${\cal M}$, $c_{+(\times)}$, $q_{+(\times)}$, 
$d^{\theta(\phi)}$ are all functions
of $(u,\theta,\phi)$. The form of these expansions follows from a careful
analysis of the characteristic initial-value problem for the vacuum Einstein
equations (see refs.\refs{\Bondi, \Sachs} and for a textbook review 
\refs{\Inverno}).

The function ${\cal M}(u,\theta,\phi)$ may be called \refs{\Inverno} 
the `mass aspect'.
Its integral over the unit 2-sphere,
\eqn\sectce{
M(u) = {1\over 4\pi} \int d\Omega_2 ~{\cal M}(u,\theta,\phi)
}
represents the mass of the system at ${\cal I}^+$ and is the familiar
Bondi mass. Similarly $d^{\theta(\phi)}$ and $q_{+(\times)}$ are 
the `dipole' and `quadrupole aspects' respectively.
${\cal M}$, $d^{\theta(\phi)}$ and $q_{+(\times)}$ satisfy dynamical 
equations derived from the Einstein equations for their $u-$derivatives,
involving also the remaining functions $c_{+(\times)}$.

$\pl_u c_{+(\times)}$ are functions which must be given as initial data
and are specified on ${\cal I}^+$. They are the Bondi `news functions'.
An especially important result relates $\pl_u {\cal M}(u)$ to the news
functions:
\eqn\sectcf{
\pl_u {\cal M}(u) = -{1\over4\pi} \int d\Omega_2~\Bigl(
(\pl_u c_+)^2 + (\pl_u c_\times)^2 \Bigr)
}
This states that the Bondi mass is reduced if the news function is 
non-zero, corresponding to the fact that the system loses mass if
and only if it is radiating.

Finally, as we discuss later, the second derivatives 
$\pl_u^2 c_{+(\times)}$ can be identified as the amplitude of the 
gravitational waves in a weak-field limit. Notice also that
$\c \pm \d$, and hence $c_{+(\times)}$ and $q_{+(\times)}$, correspond
to the two independent gravitational wave polarisations.
 
We return to the interpretation of $c_{+(\times)}$ 
in section 4 after introducing the Peeling Theorem, discussed
by Sachs in \refs{\Sachs}. This gives the leading asymptotic behaviour
in $1/r$ of the set of Newman-Penrose scalars $\Psi_0$, $\ldots$, $\Psi_4$ 
characterising the Weyl tensor in the Bondi-Sachs spacetime.
Although we do not need to exploit it here, the  $\Psi_0$, $\ldots$, 
$\Psi_4$ are intimately related to the Petrov classification, which 
classifies the Weyl tensor according to the degeneracy of its
principal null vectors. 

For our purposes, we simply need the following result (note that since
we only need the precise results for $\Psi_0$ and
$\Psi_4$ in what follows, we have only written  schematic forms for the
others):
\eqnn\sectctwo
$$\eqalignno{
\Psi_4 ~~&=~~ {1\over r}~\Bigl[
\pl_u^2(c_+ -i c_\times)
\Bigr] \cr
\Psi_3 ~~&\sim~~ {1\over r^2}~\Bigl[
{1\over \sin^2\theta} \Bigl(\pl_\theta +
{i\over\sin\theta}\pl_\phi\Bigr) \Bigl(\sin^2\theta \pl_u(
c_+ - i c_\times)\Bigr)
\Bigr] \cr
\Psi_2 ~~&\sim~~ {1\over r^3}~\Bigl[
{\cal M} + \Bigl(\pl_\theta +
{i\over\sin\theta}\pl_\phi\Bigr) \Bigl(
c_+^2 + c_\times^2\Bigr)
\Bigr] \cr
\Psi_1 ~~&\sim~~ {1\over r^4}~\Bigl[
d^\theta + i \sin\theta d^\phi
\Bigr] \cr
\Psi_0 ~~&=~~ {1\over r^5}~\Bigl[
6(q_+ -i q_\times)
\Bigr] &\sectctwo\cr }
$$
where we have set $\ell_\m = \pl_\m u$ and chosen the transverse vector
$m_\m = {1\over\sqrt2}r(\pl_\m\theta + i\sin\theta
\pl_\m\phi)$. 

The essence of the Peeling Theorem is the correlation between the leading
order in $1/r$ and the type of the NP scalar. Notice also that
the leading  coefficients as we pass from $\Psi_4$ to $\Psi_0$ involve
respectively $\pl_u^2 c$, $\pl_u c$, ${\cal M}$, $d$ and $q$, with the
higher moment aspects being associated with successively higher powers of
$1/r$.

\vfill\eject

\newsec{Photon Propagation in the Bondi-Sachs Spacetime}

We are now ready to combine the general results for modified photon
propagation in section 2 with the special features of the Bondi-Sachs
gravitational radiation spacetime. Consider first the case of photons with
momentum $k^\m = \omega \ell^\m$. This corresponds to motion radially
outwards from the gravitationally radiating source.
(To confirm this, note that the equipotential surfaces for 
outgoing waves are $\vartheta(u) = const$, so using the geometrical 
optics analysis above, the corresponding rays have tangent vector
$k_\m = \pl_\m \vartheta$, so we can identify the directions of
$k_\m$ and $\ell_\m$.) Choose the transverse polarisation vectors 
$a^\m$, $b^\m$ to lie in the $\theta$ and $\phi$ directions respectively,
so that
\eqnn\sectdone
$$\eqalignno{
a^\m &= {1\over\sqrt2}(m^\m + \bar m^\m) \cr
b^\m &= -{i\over\sqrt2}(m^\m - \bar m^\m) &\sectdone\cr }
$$
with $m^\m$ as in section 3. In this case, according to eq.\lightconenew~
the light-cone shift for the two polarisations is
\eqn\sectda{
k^2 ~~=~~\pm{4c\omega^2\over m^2}~ \bigl(\Psi_0 + \Psi_0^*\bigr)
}
which for the Bondi-Sachs metric is
\eqn\sectdb{
k^2 ~~=~~\pm {48c\omega^2\over m^2}~ {q_+\over r^5}
}
corresponding to a velocity shift $\d v$ proportional to $\pm q_+/r^5$.
As usual for Ricci-flat spacetimes, the two transverse polarisations
have equal and opposite velocity shifts, so one is always superluminal.
For outgoing photons, therefore, we do find a velocity shift, but it
is very weak, falling off as $1/r^5$, and is governed by the quadrupole
aspect $q_+$ of the gravitational field.

It is interesting also to look at the photons with polarisation
vectors rotated through $45^o$. In this case, we choose
\eqnn\sectdtwo
$$\eqalignno{
a^\m &= {1\over2}(m^\m + \bar m^\m) - {i\over2}(m^\m - \bar m^\m)\cr
b^\m &= -{1\over2}(m^\m + \bar m^\m) - {i\over2}(m^\m - \bar m^\m)
&\sectdtwo\cr }
$$
A calculation along the lines of eq.\sectbthree~ now gives
\eqn\sectdc{
k^2 ~~=~~ \pm{4c\omega^2\over m^2}~ i\bigl(\Psi_0 - \Psi_0^*\bigr)
}
For the Bondi-Sachs metric this is
\eqn\sectdd{
k^2 ~~=~~\pm {48c\omega^2\over m^2}~ {q_\times\over r^5}
}
As expected, the photons with polarisations rotated through $45^o$
are influenced by the $\times$ polarisation of the gravitational radiation,
compared with the original choice aligned with the $+$ polarisation.

A significantly larger effect is obtained if we consider incoming
photons, moving radially towards the source of gravitational radiation.
In this case, we have $k^\m = \omega n^\m$. For the initial choice of
photons polarised in the $\theta$ or $\phi$ directions, we find
\eqn\sectde{
k^2 ~~=~~ \pm{4c\omega^2\over m^2}~ \bigl(\Psi_4 + \Psi_4^*\bigr)
}
while for the polarisations rotated through $45^o$ we have
\eqn\sectdf{
k^2 ~~=~~ \pm{4c\omega^2\over m^2}~ i\bigl(\Psi_4 - \Psi_4^*\bigr)
}
In Bondi-Sachs, this means
\eqn\sectdg{
k^2 ~~=~~\pm {8c\omega^2\over m^2}~ {1\over r} \pl_u^2 c_+
}
and
\eqn\sectdgg{
k^2 ~~=~~\pm {8c\omega^2\over m^2}~ {1\over r} \pl_u^2 c_\times
}
respectively. In this case, the superluminal velocity shifts are
of $O(1/r)$, i.e.~$\d v$ is proportional to $\pl_u^2 c_{+(\times)}/r$
depending on whether the photon polarisation is aligned with the
$+$ or $\times$ gravitational radiation polarisation.

We now see clearly the relation of superluminal photon propagation 
to the Peeling Theorem. Depending on the direction (and polarisation)
of the photons, the shift in the light cone is proportional to one
of the NP scalars characterising the Weyl tensor. The 
Peeling Theorem specifies the leading order in $1/r$ of each of these 
types in the vicinity of ${\cal I}^+$, which translates immediately
into a result giving the $1/r$-dependence of the photon velocity shifts.

These results of course hold for the full gravitational radiation field
described by the Bondi-Sachs metric. It is interesting at this 
point to compare them with those obtained previously for 
weak-field gravitational radiation in the linearised approximation.
To see this relation, consider the following metric describing 
gravitational plane waves in the linearised, weak-field limit where the
metric perturbation $h_{\m\n}$ from Minkowski spacetime is chosen in
transverse, traceless gauge:
\eqn\sectdh{
ds^2 = - du dv  + \bigl(1- h_+(u)\bigr) dx^2
+ \bigl(1+ h_+(u)\bigr) dy^2 - 2h_\times(u) dx dy
}
$h_+$ and $h_\times$ of course correspond to the $+$ and $\times$
polarised gravitational waves, and $\pl_u^2 h_{+(\times)}$  
represent their amplitudes \refs{\Inverno}. 
The relevant components of the Weyl tensor are:
\eqnn\sectdthree
$$\eqalignno{
C_{uxux} = -C_{uyuy} &= -{1\over2}\pl_u^2 h_+  \cr
C_{uxuy} &= -{1\over2}\pl_u^2 h_\times  &\sectdthree\cr }
$$

Linearising the Bondi-Sachs metric in $\c$ and $\d$ and discarding the
functions $W$, $\b$ and $U^i$, we find
\eqn\sectdi{
ds^2 = -du dv + r^2 \Bigl( (1+\c+\d)d\theta^2 + 
2(\c-\d)\sin\theta d\theta d\phi + (1-\c-\d) \sin^2\theta d\phi^2 \Bigr)
}
which has therefore reduced to the weak-field gravitational wave
metric with
\eqn\sectdj{
h_+ = - (\c+\d) = - 2{c_+\over r} ~~~~~~~~~~~~~
h_\times = - (\c-\d) = - 2{c_\times\over r}
}
confirming the identification already assumed above that $c_{+(\times)}$
correspond to the two independent gravitational wave polarisations.

Returning to the metric \sectdh, the light-cone shift for photons 
travelling in the opposite direction to the gravitational waves is
\eqn\sectdk{
k^2 = \pm {8c\omega^2\over m^2} C_{uxux} = 
\pm {4c\omega^2 \over m^2} \pl_u^2 h_+  
}
for $x,y$ polarised photons, and
\eqn\sectdl{
k^2 = \pm {8c\omega^2\over m^2} C_{uxuy} = 
\pm {4c\omega^2 \over m^2} \pl_u^2 h_\times  
}
for $45^o$ rotated photons. With the weak-field identifications \sectdj,
we recover \sectdg, \sectdgg.

On the other hand, for photons travelling in the same direction as the 
gravitational waves, the light-cone shifts are proportional to the 
$C_{vxvx}$, $C_{vyvy}$ and $C_{vxvy}$ components of the Weyl tensor, 
which vanish.
For both polarisations, therefore, $k^2 = 0$.

To summarise, for weak-field gravitational plane waves, the effect
on photon velocity is as follows. Photons travelling in the same direction
as the gravitational waves feel no effect, whereas photons travelling in 
the opposite direction experience velocity shifts proportional to the
gravitational wave amplitudes $\pl_u^2 h_{+(\times)}$, depending on the
alignment of the photon and gravitational wave polarisations.

In the full Bondi-Sachs radiation metric, a similar situation holds, except 
that here the photons travelling in the same direction as the gravitational
radiation also experience a velocity shift proportional to 
$q_{+(\times)}$, though only of $O(1/r^5)$, while those travelling in 
the opposite direction experience a much larger $O(1/r)$ effect
proportional to $\pl_u^2 c_{+(\times)}$.

\vfill\eject

\newsec{Conclusion}

If the strong equivalence principle is violated, photons do not 
necessarily propagate along the null geodesics of the background curved
spacetime. The physical light cone is shifted with respect to the
geometrical one. Here, we have discussed this effect in terms of an 
effective field theory containing explicit SEP-violating interactions. 
Such terms have been shown to arise even in conventional QED in curved
spacetime through vacuum polarisation effects and are expected to appear
generically in the low-energy limit of theories of quantum gravity.

Developing earlier work on black hole spacetimes, we have considered
the special case of the gravitational radiation metric introduced
by Bondi, van der Burg and Metzner \refs{\Bondi} and by Sachs \refs{\Sachs}. 
This has allowed us
to generalise previous results on photon propagation in weak-field
gravitational wave backgrounds obtained by Drummond and Hathrell \refs{\DH}.

Our principal results are that the velocity shifts are either 
superluminal or subluminal depending on which of the transverse
photon polarisations is considered. For photons travelling inward 
towards the source of gravitational radiation, the velocity shifts are
asymptotically of $O(1/r)$ and depend on the functions 
$\pl_u^2 c_{+(\times)}$, where $\pl_u c_{+(\times)}$ are the Bondi news
functions. The light-cone shift depends on the relative alignment
of the photon and gravitational radiation polarisations. For photons
moving outwards along the direction of the gravitational radiation,
we still find a non-vanishing velocity shift (in contrast to the
case of weak-field gravitational waves) controlled by the quadrupole 
aspect $q_{+(\times)}$ of the gravitational radiation, 
but it is only of $O(1/r^5)$.

The implications of these results for the question of causality
in the presence of superluminal propagation are left for future work.

\newsec{Acknowledgments}

I am grateful to I.~Drummond, R.~Durrer, K.~Kunze, M.~Maggiore,
W.~Perkins and G.~Veneziano for helpful and interesting discussions.
This research is supported in part by PPARC grant GR/L56374.

\listrefs

\lref\DH{I.T. Drummond and S. Hathrell, Phys. Rev. D22 (1980) 343 }
\lref\Sone{R.D. Daniels and G.M. Shore, Nucl. Phys. B425 (1994) 634 }
\lref\Stwo{R.D. Daniels and G.M. Shore, Phys. Lett. B367 (1996) 75 }
\lref\Sthree{G.M. Shore, Nucl. Phys. B460 (1996) 379 }
\lref\Bondi{H. Bondi, M.G.J. van der Burg and A.W.K. Metzner, Proc. Roy. Soc.
A269 (1962) 21}
\lref\Sachs{R.K. Sachs, Proc. Roy. Soc. A270 (1962) 103 }
\lref\Done{I.T. Drummond, private communication}
\lref\Dtwo{I.T. Drummond, gr-qc/9908058} 
\lref\ABJone{J.D. Barrow and J. Magueijo, Class. Quant. Grav. 16 (1999) 1435} 
\lref\ABJtwo{A. Albrecht and J. Magueijo, Phys. Rev. D59 (1999) 043516 }
\lref\ABJthree{J. Magueijo, Phys. Rev. D62 (2000) 103521 } 
\lref\DN{A.D. Dolgov and I.D. Novikov, Phys. Lett. B442 (1998) 82 }
\lref\Ch{S. Chandresekhar, {\it The Mathematical Theory of Black Holes},
Clarendon, Oxford (1985) }
\lref\Inverno{R.A. d'Inverno, {\it Introducing Einstein's Relativity},
Clarendon, Oxford (1992) }

\bye